\documentclass[10pt]{article}

\usepackage[utf8]{inputenc}
\usepackage[numbers]{natbib}
\usepackage{csquotes}
\usepackage{newtxtext}
\usepackage{setspace}         
\usepackage{geometry}         
\usepackage{ragged2e}         
\usepackage[colorlinks=true,linkcolor=blue,citecolor=blue,urlcolor=blue]{hyperref}
\usepackage{footmisc}         
\usepackage{tabularx}
\usepackage{graphicx}  %
\usepackage{pdfpages}  

\renewcommand{\raggedright}{\justifying}

\title{\vspace{-2em}Sustainability Transitions and “Bending the Curve of Biodiversity Collapse” in the Amazon Forest}
\author{
O.Y. Romero-Goyeneche\thanks{Human Geography and Spatial Planning, Faculty of Geoscience, Utrecht University. \textit{Corresponding author:} \href{mailto:o.y.romerog@gmail.com}{o.y.romerogoyeneche@gmail.com}}
\and
M. Ramirez\thanks{SPRU - Science Policy Research Unit, University of Sussex Business School. \textit{Corresponding author:} \href{mailto:M.t@gmail.com}{Matias.Ramirez@sussex.ac.uk}}
\and
A.M. Osorio\thanks{National University of Colombia.}
\and
U. Harman-Canalle\thanks{Wyss Academy for Nature and Pontifical Catholic University of Peru.}
}
\date{}

\begin{document}

\maketitle

\begin{center}
    {\LARGE \textbf{Abstract}}
\end{center}

\justifying
\noindent

This paper undertakes an analysis of deforestation in the Amazon area using a pathways-based approach to sustainability. We ground the analysis primarily in the sustainability transitions literature but also draw a bridge with socio-ecological concepts which helps us to understand the nature of transitions in this context. The concept of a deforestation system is developed by examining the interplay of infrastructure, technologies, narratives, and institutions. Drawing on a literature review and an in-depth case study of Puerto Maldonado in Madre de Dios, Peru, the paper identifies three pathways for addressing deforestation: optimisation, natural capital, and regenerative change. We suggest that while the optimisation pathway provides partial solutions through mitigation and compensation strategies, it often reinforces extractivist logics. The study also underscores the limitations of natural capital frameworks, which tend to rely on centralised governance and market-based instruments while lacking broader social engagement. In contrast, our findings emphasise the potential of regenerative strategies rooted in local agency, community-led experimentation, and context-sensitive institutional arrangements. The paper contributes to ongoing debates on biodiversity governance by illustrating how the spatial and long-term dynamics of deforestation interact, and why inclusive, territorially grounded pathways are crucial for bending the curve of biodiversity loss.

\vspace{1em}

\vspace{1.5em}

\justifying

\section{Introduction}

Discussions on reversing deforestation in the Amazon forest and similar biospheres have traditionally focused on the socio-ecological interactions between humans and nature \citep{lapola_drivers_2023}. Building on the above notions, we propose that a pathways approach, grounded in transition studies and socio-technical systems analysis, offers valuable insights into why reducing deforestation over the long term has proven so challenging, and also points to potential ways forward. Our discussion undertakes this analysis by tracing how what we call \textit{"a system of deforestation"} was constructed in the Amazon region in the second half of the 20\textsuperscript{th}
century through a combination of new infrastructures, technologies and narratives around land use. 

The system of deforestation was defined by vast over-exploitation of natural resources, displacement of Amazonian communities, especially Indigenous communities and poorer farmers and ambiguous governance relations in areas such as property rights. This combination of factors can partly help explain the lock-in to this system of deforestation and why it remains a persistent challenge despite significant efforts to reverse the forest decline. Given this context, the paper asks what types of sustainability pathways exist in the Amazon and under what conditions might they help to break the lock-in to deforestation? Our proposition is that deforestation is not simply the result of isolated environmental degradation, but rather the outcome of a deeply institutionalised socio-technical regime. 

The analysis unfolds in three main steps. First, drawing on recent debates on pathways to sustainability transitions literature, we identify key strands in current discussions on deforestation and biodiversity conservation. Second, we trace the historical emergence of the Amazon deforestation system and the efforts to reverse its decline—through both multilateral action and grassroots movements—since the second half of the 20\textsuperscript{th} century. 

The historical analysis allows us to propose three distinct pathways for addressing the extractive logic of deforestation: the optimisation pathway, the natural capital pathway, and the regenerative change pathway. These pathways are examined by analysing the interplay of interlocking rules and routines, key actors, dominant narratives, place-specific dynamics, and infrastructure. We suggest that the optimisation pathway, which focuses on incremental innovation through mitigation and compensation strategies, will ultimately be insufficient to halt deforestation. Yet, its analyisis is useful because it helps to uncover some of deforestation’s core characteristics—particularly the interactions between natural systems, technological practices, and the lives of both resident and migrant populations. 

Finally, we assess the strengths and limitations of the natural capital and regenerative change pathways in reversing deforestation and enabling more sustainable outcomes. In the third and final section of the paper, we undertake a study of the Madre de Dios region of Peru, tracing patterns of deforestation and discuss examples of initiatives taking place in the region with communities of producers. This will assist in reappraising how the regenerative pathway is constructed. We conclude by suggesting that, given the Amazon’s unique natural and social characteristics, a more resilient pathway may lie in connecting local communities livelihood strategies with technologies and infrastructures rooted in a regenerative vision of the forest. Such a context-sensitive approach has greater potential to weaken and eventually undermine the entrenched system of deforestation. We recognize, however, that this is a challenging path—one that will require substantial investment in regenerative forest alternatives, the development of place-based institutions, and the active engagement of local actors in land use decisions. 

\section{Conceptual Foundations for Pathways in the Amazon}

Pathways analysis within the transitions literature has frequently been used as a framing device to describe transitions scenarios \citep{rosenbloom_pathways_2017}. Pathways can also help explain complex processes which include advances through the development of niches and new rules, the existence of possible rebound effects and tensions and conflicts that contribute to shape and give direction to new transition pathways. We highlight three important features of pathways analysis that we consider useful for the construction of transition pathways in the Amazon. 

The first is that pathways analysis helps describe the interlocking material and social elements that stabilize certain cognitive routines \citep{kemp_regime_1998, nill_evolutionary_2009}. These include rules and sunk costs that can constrain change processes by locking in established development trajectories. Secondly, governance issues and the tendency of powerful actors and institutions to “close down” transformation processes that benefit incumbents are also underlined \citep{smith_social-ecological_2008, stirling_direction_2009, stirling_keep_2010}. Thirdly, pathways analysis is also about identifying opportunities to weaken path dependence and lock-in \citep{rosenbloom_pathways_2017}. This can be framed by changes in the rules by which actors negotiate and interpret belief systems and guiding principles in terms of their relationships with socio-technical systems \citep{geels_typology_2007}. The deployment of more participatory approaches and contestation processes are therefore an inherent part of the process for opening up new alternative pathways. Finally, more recent research on pathways has embraced some of the complex feedback loops that can exist over time and in particular the context dependency and temporality of these pathways \citep{sovacool_how_2016}. 

Early discussions on transition pathways often adopted a typological approach \citep{grin_transitions_2010}, focusing on triggers of change such as technological substitution, optimisation, and shocks, and how radical changes can catalyse broader transformations. Optimization, in this context, is typically associated with incremental adjustments within existing socio-technical regimes \citep{geels_typology_2007}. These gradual changes aim to address contestations from outsiders opposing the negative impacts of prevailing regimes, thereby reproducing established practices and reinforcing the stability of incumbent systems. This pathway therefore often involves technological fixes and incremental innovations that allow incumbent actors to respond to external pressures—such as environmental or social contestation—without challenging the core logic of the regime. 

For \citet{geels_great_2022}, it is reconfiguration pathway in areas such as wind, solar PV and biofuels that has become dominant for decarbonisation strategies in the UK (and by extension the industrialised economies of Europe and the global north). Here endogenous interactions between multiple components or modules of a system can trigger important architectural changes in a socio-technical system (for example, battery storage and smart grids to address problems of intermittent energy from renewables). These changes may open the way for deeper changes, encourage new actors to join the existing system and create greater diversification. Nevertheless, \citet{geels_great_2022} also point out that the process tends to be led by incumbents and there tends to be little re-shaping of system architectures, such as for example through decentralisation of energy. Therefore, fundamental power imbalances are not usually addressed. 

In addition, it is important to underline that there is considerable debate over whether a reconfiguration pathway can achieve decarbonisation reductions to avoid major planetary harm. \citet{schot_deep_2018}, for example, suggest that reconfiguration alone is insufficient to account for the complexity of cross-system interactions that structure meta-regimes and unlock possibilities for change. Similar critiques from the degrowth literature share the scepticism that it is possible to decouple the environmental impact from economic growth via reconfiguration pathways, despite the deployment of renewables and the green tech revolution \citep{kallis_coevolutionary_2010}. This literature then suggests the need to rebuild the basic relationship between humans and the environment to generate sustainable pathways. These concerns offer a quite different pathway in which place-based dynamics, cross-system interactions, narratives centred on the well-being of the forest and its inhabitants (including humans), and the creation of new institutions contribute to re-establishing human–forest relations and advancing regenerative practices capable of bending the curve of biodiversity loss.

The above discussion suggests that it is possible to broadly conceptualise three types of pathways for the overcoming deforestation. The first relies on the optimisation of existing systems, where path dependence and inertia of socio-technical systems dominates. At most there are incremental changes to meet sustainability challenges \citep{turnheim_evaluating_2015}. This is partly a question of the obduracy of system architectures that lead to inertia, but regimes can also be reproduced via prevailing regulatory, normative and behavioural practices and through active defence and resistance strategies of dominant market players \citep{geels_sectoral_2004}.

Secondly, as discussed, a reconfiguration pathway in which important changes in new technologies and system components can affect broader systems changes as seen in the case of decarbonisation strategies. Finally, a deeper system change pathway where many of the basic assumptions of capitalist growth are re-thought, deeper transformations of sub-regimes take place (science and technology, industry, culture, users, and infrastructure), as well as the rules and actors of the regime. 

\section{Pathways for Bending the Curve of deforestation in the Amazon}

In this section, we focus on two primary pathways that explicitly aim to reduce deforestation: reconfiguration and what we will refer to as regenerative change. Yet, it is worth stating that optimisation-type practices continue to be favoured at different levels. For instance, improved monitoring technologies or low-impact logging techniques may help mitigate some negative effects of deforestation, but they do not transform the underlying land-use model or governance structure. In this pathway, radical niche innovations may emerge, but they are often excluded from the dominant regime’s dynamics, limiting their capacity to scale or reshape the system.

While optimisation reinforces the dominant regime through incremental adjustments, reconfiguration involves the selective integration of elements from socio-technical niches or movements that contest the regime, thereby diversifying and partially reshaping its structure and logic. In socio ecological contexts, reconfiguration can be expressed in the natural capital approach, in which natural resources are valued as a way of protecting biodiversity. Dasgupta, a well-known proponent of the natural capital approach explains that \textit{“adequately valuing”} nature so that accounting price and market price will reflect its social value is the key \citep{dasgupta_economics_2021}. In the natural capital approach, nature is treated as a financial asset, the value portfolio of which can be recognised and compared with other similar financial portfolios. By correctly pricing nature, the argument goes, it is possible to optimise resource management and channel resources—through mechanisms such as Payment for Ecosystem Services (PES)—toward nature restoration. This is achieved by incentivising the beneficiaries of ecosystem services to compensate those holding property rights (and supplying the service) for restoring, preserving, and protecting ecosystems.

The natural capital approach appears to share features of the system reconfiguration approach discussed above by \citet{geels_great_2022} in the sense that new socio-technical components of ecosystem services (modules) are developed, new actors enter to play new roles such as pricing ecosystem services and monitoring restoration of the environment plays a bigger role. Also, the architecture of environmental management becomes more decentralized. However, we also note that as a pathway, the natural capital maintains the basic rules that govern our relationship to nature, and indeed, some would argue, reinforce them. For example, the value of nature measured in individual utility and the sanctity given to the value of capital in terms of return on investment remain unchanged. Indeed, the consequence of this approach is that it can increase the dependence of small producers and Indigenous communities on market for their livelihoods. 

There are also some significant criticisms of the assumptions behind the natural capital, many of which it is claimed, bear little relationship to reality. For a more detailed critique of Dasgupta’s arguments see for example \citet{spash_dasgupta_2022}. For our purposes, the most poignant of these is the prediction by some that it will not break the continuing cycle of deforestation in areas such as the Amazon. For example, Research by \href{https://gspp.berkeley.edu/berkeley-carbon-trading-project}{UC Berkeley Carbon Trading} looking into rainforest carbon credits found that the system of carbon trading was not fit for purpose in terms of durability, forest carbon accounting, community safeguards, deforestation leakage and baselines and was open to exploitation and therefore associated schemes such as REDD+ lacked credibility. 

Dasgupta himself acknowledges, a combination of factors that might be considered as market failure and include unclear property rights, low trust in institutions of government and inability of markets to function are common in low-income countries. Consequently, this approach has shown mixed results \citep{zilberman_when_2008, pattanayak_tropical_2001}. In this regard, \citet{dasgupta_economics_2021} argues that polycentric institutions are necessary where the state, civil society and communities can construct institutions to provide public goods and services. Under this argument, the state and civil society’s role is to intervene to overcome problems of market failure. 

There is a third pathway, which assumes that conditions of market failure are the rule rather than the exception and may therefore be particularly relevant for areas such as the Amazon. This pathway does not rely on market incentives but is instead grounded in alternative value systems that shape the relationship between humans and nature \citep{brondizio_making_2021}. It includes a broad range of imaginaries, narratives, practices, and social movements such as the \textit{“rights of nature”}, Indigenous ancestral agriculture \citep{rudel_indigenous-driven_2021}, various strands of agroecology including syntropic and permaculture approaches, and regenerative value chains. These initiatives can be grouped under socio-ecological approaches which begin with a commitment to regenerating and enriching the forest while allowing for responsible forms of extraction \citep{szlafsztein_development_2014}. 

The regenerative pathway also emphasizes the importance of place by acknowledging variations in climate, soil, and ecological conditions. Regenerative strategies therefore require careful stewardship and the active involvement of communities whose livelihoods are closely tied to these practices. In many cases, this involves the engagement of both Indigenous and non-Indigenous groups, whose attachment to the land contributes to the strengthening of commons institutions and supporting infrastructures, such as seed banks, ecosystem-based production systems, and intercropped agriculture \citep{rudel_indigenous-driven_2021}.

A key element of this approach is the emergence of new spatial imaginaries rooted in lived experience and the reintegration of territories into ecological cycles. These imaginaries enable connections between different producer communities, including those involved in agriculture, mining, and agroforestry. Narratives such as \textit{“the health of the forest is the well-being of its people and the land”} help guide the formation of alternative systems through experimentation with technologies, practices, and land uses. In what follows, we examine the evolution of the Amazon forest as a socio-technical system and explore how these pathways have materialised in practice.

\section{Methods}

The discussion so far has outlined a possible pathways approach that distinguishes between different strategies for combating deforestation. In the next sections, we analyse changing patterns of land use in the Amazon from the second half of the 20\textsuperscript{th} century to further understand how these pathways may be relevant to current debates on deforestation. This relies on work on the Amazon from the period when a change in land use patterns accelerated in the 1950s and on a set of interviews in the Madre de Dios region of the Peruvian Amazon. We draw on partly on \citet{geels_sectoral_2004, geels_great_2022} that make reference to three main elements of a socio technical systems; techno-economic components and flows (material artefacts, goods, infrastructures, consumption patterns). 

Secondly, the actors and their organisation into social groups such as business, policymakers, uses and social groups, each of which will have domains of practice, arguments and levels of agency the rules (regime or niche) that influence how problems are framed. Finally, there are policies that are guided by rules and institutions and can include the degree of policy coherence, consistency and comprehensiveness of instruments.

We build on these categorisations by using two adapted analytical dimensions, namely socio-economic infrastructures underpinning forest management and use and secondly, actors and their governance structures and socio-technical rules expressed through dominant land-use narratives. To this we another dimension, which is place-based analysis that we link to socio-ecological dynamics. Underpinning the above is a relational approach to governance, understood as socially constructed spaces that frame the interactions between niches, regimes and landscapes \citep{pachoud_societal_2022}

\subsection{Data}

The literature review was conducted using the Web of Science Core Collection, focusing on peer-reviewed journal articles and book chapters related to the history of deforestation in the Amazon. The search query used was:\\ 

\textit{TS=(Amazon AND deforestation OR biodiversity loss AND history)}.\\

From an initial pool of over two hundred results, seventy articles were preselected based on a review of their abstracts. A smaller subset was then selected by prioritising publications from the last decade in order to capture both the historical evolution of deforestation and recent analytical trends that inform the development of the pathways. 

In total, thirty-six articles were selected and read in full. The majority of the case studies discussed in these works focused on Brazil. We acknowledge that this may introduce a certain bias in our review; however, relevant scholarly studies showing similar pathways in Bolivia \citep{killeen_total_2008, mckay_agrarian_2017, millington_scale_2003, steininger_tropical_2001}, Peru \citep{arce-nazario_human_2007, da_conceicao_redd_2018}, 
Ecuador \citep{erazo_landscape_2011, finer_ecuadors_2009, messina_land_2006, rudel_indigenous-driven_2021, viteri-salazar_expansion_2020}, and Colombia \citep{armenteras_patterns_2006, etter_historical_2008, negret_emerging_2019, rodriguez_patterns_2012} are also considered.

Secondly, in Section 6, we present an in-depth case study of the Amazonian region of Tambopata in Madre de Dios, Peru—an area profoundly shaped by the dynamics of the deforestation system, yet also marked by significant local efforts aimed at reversing it. Located at the confluence of the Tambopata and Madre de Dios rivers, the city has undergone rapid socio-environmental transformation, particularly since the construction of the Interoceanic Highway, which has intensified access and catalysed deforestation. The region’s high ecological value, exemplified by protected areas such as the Tambopata National Reserve, makes it a critical site for analysing sustainability pathways. 

The study draws on a robust and integrated methodological framework, combining a series of workshops and semi-structured interviews with key stakeholders conducted during multiple field visits between August 2023 and April 2025. The workshops were organised by the Wyss Academy for Nature’s Latin America Hub and additional interviews were conducted with actors who were unable to attend. The authors also carried out extended fieldwork in the region over several weeks, which included site visits, informal interactions, and the review of relevant policy documents and project reports. This multi-method approach enabled the triangulation and validation of findings by integrating narratives from grassroots leaders, institutional stakeholders, and policymakers. Taken together, these sources provide a comprehensive empirical foundation for analysing how alternative pathways to deforestation are being envisioned and enacted in practice.

\section{The Amazon’s system of Deforestation}

This section provides a discussion of the conditions that led to the emergence of the \textit{"system of deforestation"} in the Amazon rainforest, followed by an exploration of four-time phases that are described below: branching, natural capital, regenerative change and rebound effect.

\subsection{The Emergence Phase}

Prior to the mid-20\textsuperscript{th} century, the Amazon was primarily a regenerative system that brought significant natural benefits to humanity and all living beings (carbon capture, regulation of the climate among other ecological functions of the forest). However, from the second half of the 20\textsuperscript{th} century, \citet{russo_lopes_understanding_2022} argue that a regime of deforestation emerged in the Amazon which was built on principles of commercialising forest products such as logging, wildlife trade, intensive agriculture, hydro-electric dams, intensive agriculture and beef production to supply the growing demand in urban areas for housing, food, animal feed, cheap energy and mobility \citep{rudel_forest_2005, prates-clark_implications_2009, da_silva_political_2022}

The \textit{‘system of deforestation’} emerged during Brazil's military dictatorship (1965–1985), driven by land reforms aimed at regional development (Table 1). These reforms portrayed the Amazon as an underutilised frontier urgently requiring economic exploitation \citep{brondizio_making_2021, brondizio_level-dependent_2012, fearnside_deforestation_2005, viteri-salazar_expansion_2020}. The Brazilian military regime played a pivotal role in accelerating large-scale deforestation through initiatives such as \textit{‘Operation Amazonia’"}, which actively promoted the migration of landless peasants into the Amazon region \citep{kohler_local_2015, simmons_amazon_2007, simmons_political_2004}

Centralised governance institutions, notably the Superintendence of the Amazon (SUDAM), facilitated the rapid colonisation of these new settlers, enabling extensive resource extraction through infrastructure development and financial incentives tailored for agribusiness expansion \citep{fearnside_deforestation_2005}. The colonisation severely marginalised Indigenous communities and local smallholders, including rubber workers, economically pressuring them and driving their migration toward newly emerging deforestation frontiers \citep{aldrich_contentious_2012, kohler_local_2015}, thus exacerbating ecological degradation. These marginalised groups organised unions and actively contested the system of deforestation, connecting their claims to emerging environmentalist NGOs and the growing global concern regarding environmental impacts \citep{brondizio_making_2021}, as will be outlined in more detail below. 

Socio-technical rules underpinning the system of deforestation were closely tied to the narrative of the Amazon as an economically \textit{"unproductive"} region \citep{schmink_contested_2019} urgently requiring pioneering colonisation and economic modernisation. \citet{kohler_local_2015} clearly illustrates the narratives surrounding the necessity of deforestation for economic development. For example, the slogan \textit{"land without people for people without land"} vividly encapsulated this narrative, legitimising extensive deforestation and environmental degradation by presenting large-scale agriculture as a development strategy. Similarly, agricultural modernisation was portrayed as a way to overcome \textit{"peasanthood"}, representing a new social interaction with the forest based on exploitation. Additionally, colonisation efforts began through the official state concept of \textit{"vacant lands"}, implying that lands already inhabited were considered insufficiently utilised and, therefore, eligible for conversion and productivity.

\begin{center}
    \textbf{Table 1.} Pathways in the Amazon Rainforest
    
    \vspace{0.5em}
    \includegraphics[width=\textwidth]{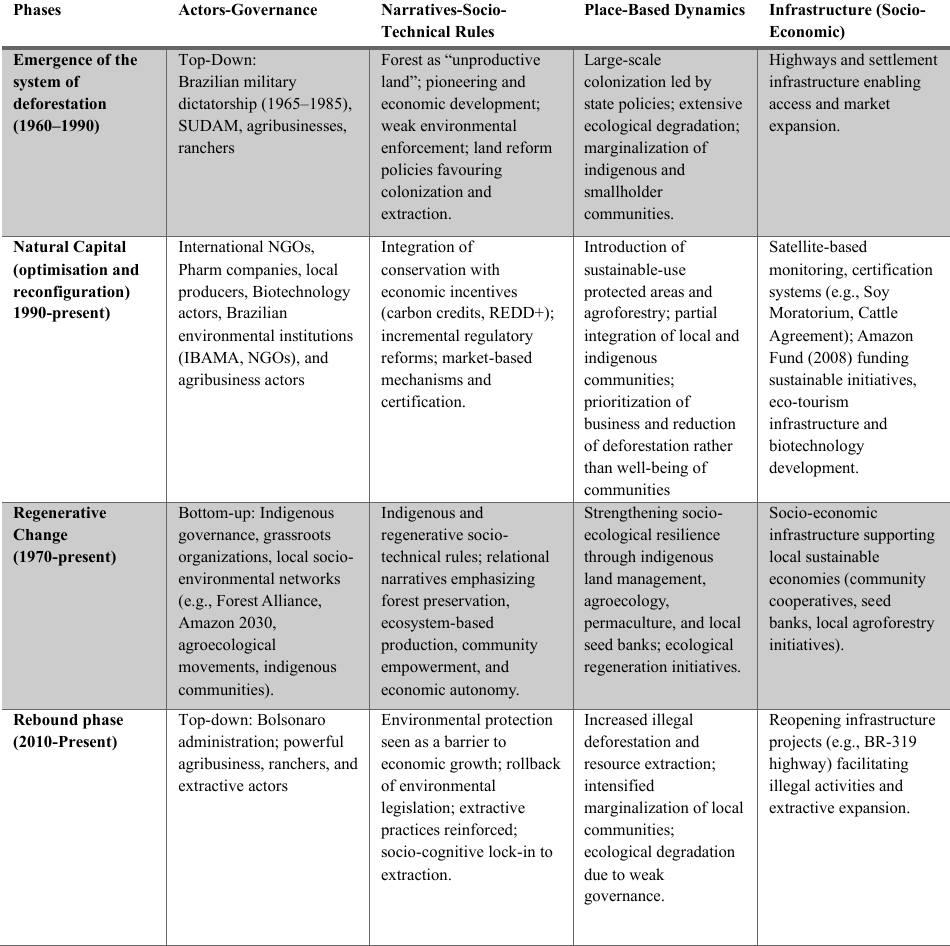}
\end{center}


\justifying

Reinforcing these dominant narratives were key regulations that institutionalised land-use transformations, notably as the case of Ecuador shows -the Agrarian Reform and Colonization Acts enacted in 1964, 1973, and 1979 as presented by \citet{viteri-salazar_expansion_2020}. Specifically, the first Agrarian Reform Act of 1964 aimed at dismantling the Indigenous family-based, huasipungo system, breaking up large lands historically controlled by wealthy landowners and the Catholic Church, and reallocating these lands to peasants. Subsequently, the second Agrarian Reform Law (Decree No. 1172, 1973) pursued redistribution of land to improve the living conditions of new agricultural producers and further promoted settlement in previously forested regions. This colonisation process was intensified by additional legislation, such as the Law of Agricultural Promotion and Development (Supreme Decree No. 3289, 1979), providing subsidies and compensating beneficiaries within traditional property structures. Moreover, the third Agrarian Reform and Colonization Act of 1979 strategically encouraged colonisation predominantly by non-Indigenous groups from other regions, explicitly targeting border areas for geopolitical as well as economic purposes. 

Collectively, the three land reforms institutionalised the system of deforestation, significantly accelerating monoculture expansion and deforestation. The rules and narratives established during this period created enduring conditions that imprinted systemic patterns of large-scale ecological degradation and socio-economic inequalities, promoting clusters of industrial production interconnected with global markets \citep{cerri_recent_2005, wassenaar_projecting_2007}. 

In particular, infrastructure developments such as the Trans-Amazonian Highway \citep{fearnside_deforestation_2005}, constructed during the 1970s and 1980s, dramatically accelerated ecological transformations by providing access to previously remote forested areas, greatly intensifying deforestation and giving access to goods produced in the Amazon to the global market. Simultaneously, the introduction of large-scale agricultural technologies designed to improve soil quality facilitated expansive monoculture production of crops such as soy and corn, which were closely linked to the industrialisation of cattle ranching \citep{steward_colonization_2007}. Hence, the infrastructure of this period was intentionally designed to maximise resource extraction from the Amazon, integrating these industrial clusters into global supply chains. 

\subsection{The Branching Phase}

By the late 1980s, international NGOs, local resistance groups and global environmental movements contested the extractivist model. This new political agenda embodied in sustainability created what Polanyi describes as a “double movement”: on one side, market-oriented mechanisms to install a more self-regulated system to allow continued exploitation of the forest, and on the other, initiatives promoting social well-being towards conservation and restoration of the forest. 

Consequently, two distinct but interrelated pathways emerged. The first involved the optimisation and reconfiguration of the extractivist model toward green industrialisation of the forest, which has parallels with Dasgupta’s concept of natural-based capital (Table 1). Here, the dominant narrative asserted that industrial forest exploitation could be more beneficial than the grassroots use of the forest, as in the case of rubber or nuts collectors. The second pathway was anchored in grassroots movements advocating for a locally embedded model of forest use, deeply integrated with the forest’s socio-ecological dynamics. In many ways, this linked up and strengthened previous ancestral practices used by indigenous populations in the Amazon that depended on the health of the forest for sustained living. In the next section, we elaborate further on these two pathways in greater detail.

\subsection{Natural Capital phase: Optimisation and Reconfiguration of the System of Deforestation}

The narratives and rules underlying the natural capital pathway revolve around the marketisation of the forest through various mechanisms \citep{dasgupta_economic_2024, dasgupta_economics_2021}. In the case of the Amazon, the natural capital pathway (Table 1) is supported by Carbon Credits like REDD+, established financial mechanisms to support forest conservation, along with sustainable practices and initiatives such as eco-tourism, bioeconomy and other strategies that promote forest sustainability.

In addition, market-oriented agreements such as the Soy Moratorium (2006) significantly shaped the natural capital narrative through the optimisation or mitigation of industries tied to deforestation. Signed by Brazil's two largest soy producer associations, covering approximately 90\%
of the Brazilian Amazon soy market, this moratorium restricted purchases from areas deforested after July 2006 \citep{garrett_forests_2021}. 

Similarly, the G4 Cattle Agreement, later expanded into the G6 Agreement, was established through a partnership between Greenpeace and major cattle companies, mandating deforestation monitoring among the use of land for cattle and banning ranchers who deforested after 2009 \citep{garrett_forests_2021}. Complementing these measures, the Brazilian Agricultural Research Corporation (EMBRAPA) introduced the Carbon Neutral Beef initiative in 2015 to certify beef produced under sustainable systems, while organisations such as the Rainforest Alliance and Round Table in Responsible Soy (RTRS) offered certifications for sustainable producers, facilitating preferential market access \citep{garrett_forests_2021}.

Additionally, the natural capital pathway has been supported by a global regulatory landscape addressing climate change and biodiversity loss. Between the 1990s and 2010s, forest management in the Amazon was significantly influenced by reduced state involvement, deeper integration into global markets, the flexibilization of labour practices, and the strengthening of international environmental regulations \citep{brondizio_making_2021}. Brazil's commitment to environmental regulations was marked by signing the Kyoto Protocol in 2002, which involved national greenhouse gas emission reduction targets and introduced restoration initiatives financed by the \href{https://unfccc.int/process-and-meetings/the-kyoto-protocol/mechanisms-under-the-kyoto-protocol/the-clean-development-mechanism}{Clean Development Mechanism} to finance the reduction of greenhouse emissions. 

Later, commitments under the 2015 Paris Agreement and Sustainable Development Goals (SDGs) further reinforced national strategies to reduce deforestation and greenhouse gas emissions. Domestically, Brazil launched the Action Plan for the Prevention and Control of Deforestation in the Legal Amazon (PPCDAm) in 2004 under President Lula’s administration, resulting in a 80\% reduction in deforestation rates by 2012 \citep{brondizio_making_2021}. This program expanded conservation units by 20 million hectares and added 10 million hectares to protected areas, alongside stricter enforcement against forest clearing on private lands, which contain nearly 60\% of Brazil’s remaining native vegetation \citep{garrett_forests_2021}. 

Our assessment of the natural capital pathways in the Amazon region, therefore, suggests that the main narrative is that sustainable industrialization of the forest provides more significant long-term benefits for the country's economy and peasants in the Amazon region. This can be achieved by optimising existing unsustainable industries, such as soybean and cattle production, while also reconfiguring the system of deforestation by diversifying the use of the forest through more sustainable practices such as bioeconomy and ecotourism, but without fully including local communities in policy and decision-making. 

There are important questions regarding this pathway. The first is that the natural capital pathway seems to have triggered disparities in the region, highlighting complexities and limitations within this pathway. Although initiatives such as REDD+, the Soy Moratorium, and certification schemes contributed to notable reductions in deforestation rates, they frequently preserved the underlying extractivist logic rather than fostering genuine systemic transformation \citep{garrett_forests_2021}.

Some authors \citep{da_conceicao_redd_2018, da_silva_political_2022} critique jurisdictional REDD+ initiatives for reinforcing historical colonial dynamics, noting that these programs disproportionately benefit powerful stakeholders, continuing the marginalising local and Indigenous communities. Similarly, the Soy Moratorium, despite its notable success in temporarily reducing deforestation, largely strengthened industrial agriculture by collaborating with major agribusiness corporations \citep{steward_colonization_2007}, thus maintaining existing power structures rather than transforming them. 

\citet{schmink_contested_2019} provide further evidence of the limitation of the natural capital pathway, arguing that Amazonian \textit{‘development’} narratives shifted historically from extractivist to agrarian colonialist (e.g., soy and cattle) and subsequently to green-environmentalist models, yet continuously favoured large-scale use of the forest. In addition, powerful stakeholders and large landowners demonstrated a high capacity to adapt to evolving frontier conditions, easily exploiting these transitions—whether extractive, agrarian, or environmentally \textit{"green"}. Conversely, less influential stakeholders were frequently excluded from benefiting from these shifts. Hence, the transition toward natural capital pathways has achieved some positive environmental outcomes, reducing deforestation for some periods, but might have simultaneously reinforced existing socio-economic inequalities, perpetuating historical patterns of exclusion by favouring larger, more resourceful stakeholders over smaller local actors and Indigenous communities. 

There is also the question over whether natural capital practices that underpin the reconfiguration pathway demonstrate sufficient resilience in reducing deforestation and restoring biodiversity? This pathway is highly dependent on institutional frameworks and governance agreements which must be operationalised by practitioners. However, this dependency presents considerable challenges in regions such as the Amazon, where property rights are ambiguous, institutional performance is weak, and common-pool resource governance is regionally highly uneven. Consequently, more durable outcomes are often restricted to some peri-urban areas, where monitoring and enforcement are more practicable. Beyond these areas, progress is frequently ephemeral, particularly in the absence of consistent financial incentives for forest conservation or where monitoring mechanisms are difficult to implement (see, for example, \href{https://www.theguardian.com/environment/2025/apr/17/revealed-worlds-largest-meat-company-jbs-may-break-amazon-deforestation-pledges-again}{World’s largest meat company may break Amazon deforestation pledges again}). In many cases, deforestation may be displaced to regions with fewer regulatory constraints or limited enforcement capacity.

A more fundamental critique that might explain the weak resilience of the natural capital pathway in contexts such as these is the inadequate account it provides for the substantial ecological, social, and institutional heterogeneity that characterises the Amazon basin and areas like it and its a-spatial character. Critical biophysical variations—such as climate conditions, soil fertility, water availability, and the presence of extractive resources—substantially influence the strategic decisions of local agricultural producers yet are typically overlooked. 

Additionally, the levels of social cohesion and social capital, which are pivotal in shaping collective environmental behaviour, receive insufficient analytical attention. However, in the absence of robust institutions and widespread trust in national governance structures, the realisation of sustainability objectives will depend more heavily on the degree of community alignment with environmental goals. This reliance represents a significant constraint for top-down interventions.

\subsection{Rebound Effect}

Evidence of the questionable resilience of the natural capital pathway emerges from what occurred under the Bolsonario government (Table 1), that marked a significant rebound of the \textit{‘system of deforestation’}, reversing prior progress by weakening governance, dismantling regulatory frameworks, and reinforcing extractivist interests \citep{pelicice_political_2021}. Institutional dismantling of environmental agencies such as Brazilian Institute of Environment and Renewable Natural Resource (IBAMA) and Chico Mendes Institute for Biodiversity Conservation (ICMBio), coupled with the systematic weakening of the Forest Code, exemplified this governance regression, enabling intensified resource extraction, accelerating deforestation in the region \citep{brondizio_making_2021}. Infrastructure projects, notably the reopening of the BR-319 highway \citep{ferrante_brazils_2021}, further exacerbated illegal logging, mining, and land grabbing, deepening the marginalisation of Indigenous communities and significantly increasing ecological degradation. 

During this rebound phase, socio-technical narratives reverted to earlier extractivist use of the forest, portraying environmental protections as barriers to economic prosperity. \citet{pelicice_political_2021} clearly articulate this narrative shift, highlighting two primary mechanisms: regulative actions, such as budget cuts, replacement of environmental officials, denial of Indigenous territories, and opening protected areas to extractive production and adopting the narrative of the Amazon region as a homogeneous area lacking social and ecological diversity, thus justifying infrastructure developments as pathways to deliver essential services to underdeveloped areas. 

Economic incentives once again prioritised resource extraction over sustainability, reinforcing historical inequalities and the continued marginalisation of vulnerable populations, including smallholders and Indigenous communities. Actors driving this resurgence primarily included rural agribusiness elites, loggers, miners, and other powerful economic stakeholders whose interests aligned with the extractivist paradigm \citep{brondizio_making_2021}. Consequently, the resurgence of extractivist socio-technical logics reinforced longstanding patterns of marginalisation, displacement, and environmental degradation, intensifying socio-economic inequalities and ecological damage \citep{ferrante_brazils_2021, pelicice_political_2021}. 

\subsection{The Regenerative Pathway}

The regenerative pathway in the Amazon reflects different elements that have in common a view of the Amazon that is based on economy activity that is ecologically regenerative, promotes economic activity that works in harmony with the environment, promotes community wellbeing but also its involvement in monitoring and stewarding the wellbeing of the forest. This has historically been expressed in different ways. From a civil society perspective, grassroots social movements in the Amazon have traditionally played an important role in contesting extractive pathways and mobilizing for ecologically sustainable forms of living in the Amazon. The most well- known case followed the murder of Chico Mendes, a human rights and ecological leader in the Amazon region in 1988, that escalated the contestation against the system of deforestation \citep{brondizio_making_2021}. This movement gained momentum following the 1992 Rio Earth Summit, a pivotal event that reshaped land-use narratives by legitimising environmental conservation and sustainability alongside economic development. Social movements, therefore, are an integral part of the institutional landscape of the Amazon. 

It is also important to note that social movements overlap with grassroots place-based socio-environmentalist movements integrating advocacy for local and Indigenous rights with environmental conservation community empowerment by strengthening social capital and reinforcing community identities \citep{szlafsztein_development_2014}. This approach has fostered a range of sustainable place-based alternatives to industrial exploitation, including agroecology, permaculture, agrobiodiversity, and community-driven forest management \citep{garrett_forests_2021, oestreicher_livelihood_2014}. \citet{brondizio_making_2021}, which underscores the importance of the concept of \textit{"community"} as a powerful narrative facilitating socio-ecological resilience, shared identities, and collective management of commons essential for sustained transformation. The infrastructure developed by these regenerative pathways emphasised the use of communal spaces, positioning the forest and its ecological dynamics as fundamental components within local production processes. 

However, despite their potential, grassroots movements have struggled to scale due to significant resource constraints and institutional barriers to the implementation of bottom-up initiatives in the region \citep{schwartzman_natural_2013}. One explanation presented in the literature suggests that interactions between regenerative alternatives and natural capital may have created more barriers than opportunities. Their interaction sometimes resulted in socio-technical lock-ins for grassroots actors, as powerful industrial stakeholders often appropriated emerging economic opportunities, creating dependencies within local communities \citep{da_silva_political_2022} towards monoculture, for example local production of cocoa dependent on the global supply chain of this product \citep{salazar_challenges_2023}. Despite the purportedly participatory processes involved in implementing sustainability measures, municipal and local governments often lacked the necessary resources to enforce these transitions or support smaller stakeholders effectively. As noted, \textit{“in the absence of these resources, only the more capitalised producers were able to invest in their properties, taking advantage of the new policy moment to further their claims to land and burnish their green credentials”} \citep{schmink_contested_2019}.

\subsection{From Disruption to Regeneration}

This subsection synthesis the phases underpinning the emergence and development of the deforestation system in the Amazon, as well as the alternative pathways that have arisen in response to challenge and transform it. \textbf{The emergence of the system of deforestation} in the Amazon is characterised by a radical, abrupt and imposed disruption of the relationship between local communities and their environment. This transformation was initiated through authoritarian, top-down governance structures that sought to fundamentally reconfigure land use. These transitions were not only rapid and enforced but also accompanied by systemic violence, resulting in the imposition of a new land-use regime that disregarded local socio-ecological relations and territorial knowledge.

In response to growing discontent with these authoritarian land transformations, the deforestation regime introduced strategies aimed at mitigating or compensating for negative impacts. These responses primarily focused on technical and managerial solutions—such as sustainable intensification, offsetting schemes and conservation reserves. However, such strategies did not challenge the underlying logics of extraction, centralised control and developmentalist imperatives. As these strategies proved inadequate in addressing the structural drivers of deforestation, social contestation intensified. A diversification of responses began to emerge, giving rise to alternative visions and practices (\textbf{branching phase)} that transcended technocratic approaches. Within this context, two prominent pathways gained traction: the natural capital pathway and the regenerative change pathway.

\textbf{The natural capital pathway} emerged in settings marked by institutional disorganisation, socio-political fragmentation, and escalating violence. Governance was reoriented around market-based mechanisms, such as cost-benefit analysis and valuation of ecosystem services. While this approach introduced a degree of policy and institutional diversification, it remained fundamentally centralised, with decision-making concentrated among experts and corporate actors whose primary objective was the commodification of forest resources. Land was thus primarily assessed through its economic utility, marginalising its ecological, cultural, and symbolic dimensions.

Nevertheless, market-based mechanisms have consistently failed to establish the institutional foundations required for long-term sustainability or for enabling transformative niches. This failure has led to a \textbf{rebound effect}, whereby the persistence of illegality, socio-economic inequality, and governance deficits further entrenched deforestation dynamics. National governments frequently aligned with extractive interests, privileging short-term economic returns over ecological integrity and social justice. In such contexts, local actors continued to be excluded, and violence was often employed to suppress community mobilisation and undermine alternative territorial arrangements.

As an alternative to deforestation and the depletion of natural capital, the \textbf{regenerative change pathway} offers a different approach rooted in the reconstruction of socio-ecological relations based on local knowledge systems, community engagement, and long-term environmental stewardship. It seeks to re-establish the functional and symbolic relationships between local communities and their territories through participatory governance, ecological restoration, and cultural revitalisation. This pathway recognises the cumulative degradation of ecological systems and emphasises the need to restore ecological functionality as a basis for enabling sustainable and locally appropriate development trajectories. 

As a response to the systemic socio-ecological degradation driven by the deforestation regime, local actors have mobilised to reconstitute the functionality of forest landscapes. This pathway is grounded in place-based dynamics, where communities draw upon deep ecological knowledge, affective attachments to place and collective responsibility. Regeneration thus becomes a multidimensional project—ecological, social, and institutional—focused on re-establishing viable conditions for local agency and resilience. However, the institutionalisation of these alternatives has remained limited due to persistent economic and political constraints. The recent resurgence of deforestation dynamics underscores the fragility of both natural capital reforms and grassroots initiatives, highlighting the urgent need for systemic transformation capable of addressing deep-rooted inequalities and overcoming prevailing socio-technical lock-ins.

\section{Redefining Spaces: The Case of Madre de Dios, Peru}

This section looks at a series of producer initiatives in one region of the Amazon and investigates how different actors and collectives develop strategies that address the challenges of deforestation. These initiatives relate primarily to producer practices, but the analysis is set in a particular regional context. Place and the strategies of actors to redefine it will represent a critical feature in consolidating regenerative change practices. The discussion is set in the Madre de Dios (MdD), region of Peru, an area where deforestation is taking place and accelerating (Figure 1). 

The MdD region covers 12\% of the Peruvian Amazon and is where researchers involved in this project had access to primary and secondary information on the policies and practices associated to face deforestation. The first part of the discussion is based on an analysis of official documents. We then report on the strategies of different producers based on semi-structured interviews and finally we broaden out again to the region to discuss how different commons institutions are being established and how this relates to the broader discussion of deforestation in the region. 

\subsection{Background Madre de Dios, Peru}

We begin by providing some descriptive background to the growing deforestation challenge in Madre de Dios, Peru. This region is located in the Southeast of Peru, on the border with Bolivia and Brazil (Figure 1a-b). Deforestation has increased in the three provinces that comprise Madre de Dios (Tambopata, Manu, and Tahuamanu) between 2001 and 2023, as shown in Figure 1. A commonly reported driver of deforestation in the literature is the construction and opening of the Interoceanic Highway between 2009 and 2012. As also reflected in historical analyses, the highway increased the viability of deforestation-related activities along its route, raised land prices, and contributed to the displacement of local communities. 

Additionally, the policy document \href{https://www.gob.pe/institucion/regionamazonas/informes-publicaciones/3119556-estrategia-regional-de-desarrollo-rural-bajo-en-emisiones-erdrbe}{Regional Strategy for Low-Emission Rural Development (2022)}
indicates that the area's corn cultivation quadrupled in size between 2013 and 2016. An additional interesting feature highlighted by the same document is that during the last 19 years, the loss of forest cover has been greatest in units of less than 5 hectares, although the biggest increase since 2017 has been in units of 5-50 ha. Finally, the same document concludes that the problems of deforestation emerge mainly from the occupation of lands linked to gangs that operate illegal mining and logging. This involves farmers’ associations invading or taking over lands that, over time, pass over to the hands of illegal miners once the forest has already been cleared of trees. Factors such as the illegal traffic of land ownership, the lack of monitoring and control all help to maintain high levels of deforestation. 

Several region-wide initiatives have been led by \href{https://produccionsostenible.org.pex/en/}{The Coalition for Sustainable Production}—a multi-stakeholder platform in Peru composed of public, private, and civil society organizations—to promote the transition toward deforestation-free production. The coalition aim to \textit{“increase productivity without expanding the agricultural frontier”} and prioritize value chains that protect forests and restore already degraded soils. According to its documents, for a product to be considered \textit{"deforestation-free"}, it must demonstrate, through traceability, that it was not grown on land that was deforested after 31 December 2020, the reference date established by the Coalition. This requirement is aligned with international regulations such as the The EU Deforestation Regulation, which restricts access to markets such as the European market for products linked to deforestation after that date. 

When introduced as part of a diverse agroforestry strategy, these natural capital-type initiatives can have overall positive results in terms of recovering degraded land and conservation of forests and are important steps towards significantly changing producer practices, the effects of which may not filter through to lower deforestation immediately but can be significant. One possible limitation is that they are usually tied to certain value chains, such as the European export-led value chains related to Green New Deal, the main international market for Peruvian coffee. Outside of this, Coffee and Cacao can be grown as a monoculture for national markets or other less stringent international markets. Growing coffee or Cacao as a monoculture will have negative impacts on biodiversity by simplifying landscapes, creating excessive pressure on water supplies and displacing more comprehensive approaches such as diversified agroforestry.


\begin{center}
    \includegraphics[width=0.92\textwidth]{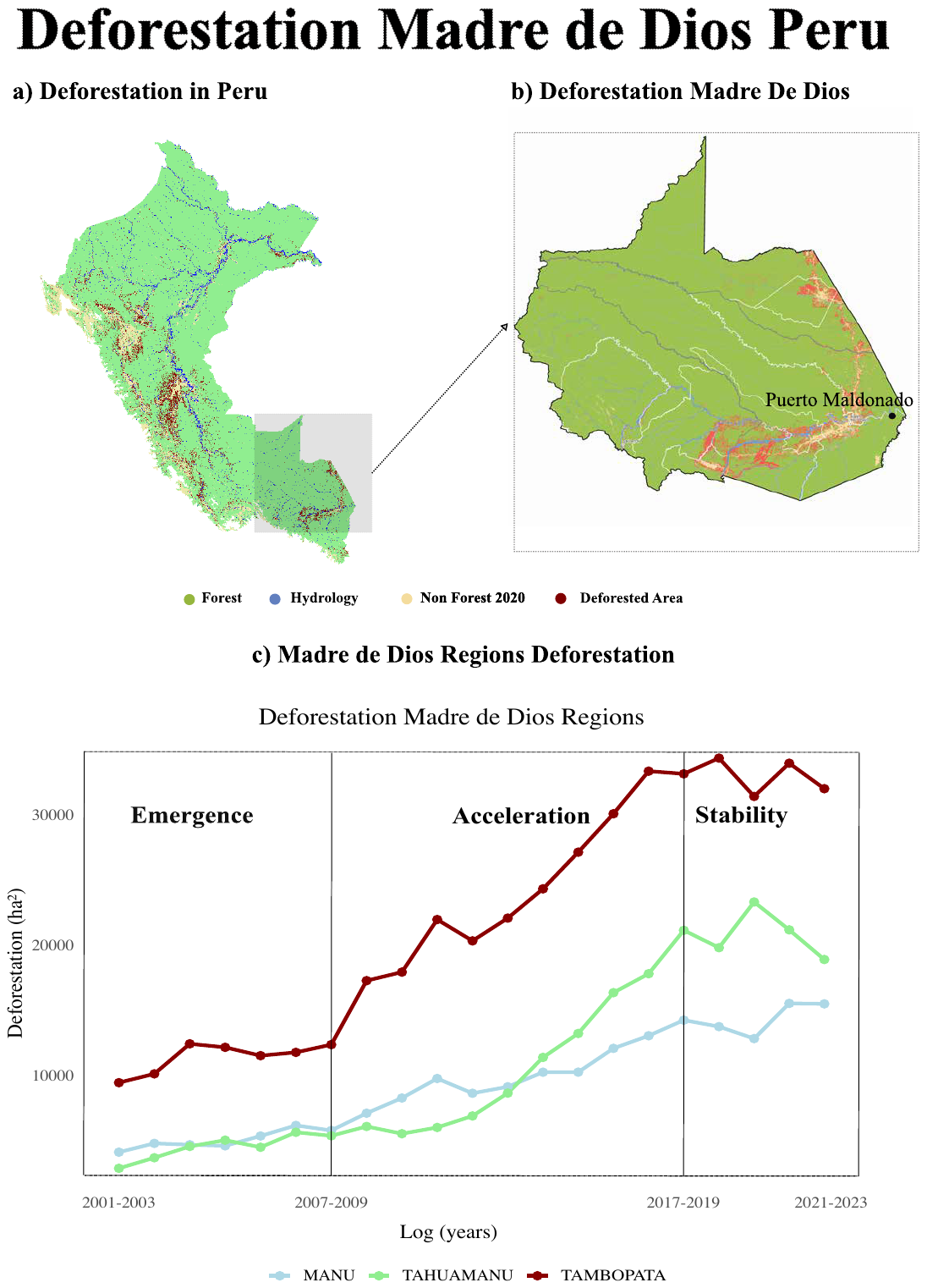}
    
    \vspace{0.5em}

    \begin{minipage}{0.92\textwidth}
    \raggedright
    \textbf{Figure 1}. Madre de Dios, Peru. (A) Deforestation in Peru. (B) Deforestation in Madre de Dios. (C) Deforestation in the three provinces of Madre de Dios from 2001 to 2023 – Rolling Window. 
    \textit{Note: Adapted from} \href{https://geobosques.minam.gob.pe/geobosque/view/descargas.php?122345gxxe345w34gg}{GeoBosques}.
    \end{minipage}
\end{center}


\justifying

Another major challenge with the Cacao and Coffee example is the implications of successfully scaling a programme such as this. At this point, these are experiments taking place in very specific locations with a plan to roll out at the national level. However, this expansion would require the ability to carefully coordinate implementation of policies and practices across privately held lands, cooperatives, small, medium and large-sized producers and to cover a vast territory well beyond peri-urban areas. However, poor monitoring capability, the opportunity costs related to illegal mining and general institutional weakness are endemic problems, especially away from urban centres. Therefore, the effectiveness of initiatives that rely only on individual monetary incentives without strong institutional or community support can be questioned.

Below we discuss in more detail three cases in the MdD region that are more akin to place-based dynamics in that they consciously introduce forest regenerative practices. Our interest is to identify how more radical changes in land use relate to the livelihoods of the actors involved, the resilience of these initiatives and other institutional features that may be important. In this sense it is important to underline that the three cases involve producer associations or individuals that are networked through funding from regional or other multilateral institutions.

\subsection{Artisan Women Gold Miners (MAPE)}

The first example is a collective of artisan women gold miners (MAPE) who have been awarded mining concessions for small-scale production that are tied to a series of practices that reduce the environmental damage normally caused by gold mining activity in terms of deforestation and water contamination. These initiatives are usually located in early stages of ecological successions where the forest and soil have been deeply affected. MAPE has an advisory board that includes representatives of the Ministry of Energy and Mines (MINEM), the Centro de Innovación Científica Amazónica - CINCIA, Cáritas, Alianza por una Minería Responsable - ARM, Pure Earth and USAID. 

An important point to underline here is that although mining has been identified as one of the main drivers of deforestation \href{https://www.gob.pe/institucion/regionamazonas/informes-publicaciones/3119556-estrategia-regional-de-desarrollo-rural-bajo-en-emisiones-erdrbe}{Regional Strategy for Low-Emission Rural Development (2022)}, all actors interviewed for this research, including agriculturalists, recognise the importance and legitimacy of \textit{“legal”} mining in the region, when conducted through legal formalised practices and its importance in providing employment and financial resources to the city through taxes.

By contrast illegal mining is frowned on because of the damage it causes, the poor labour practices and the fact that it does not contribute at all to the region. \href{https://planetforward.org/story/mercury-free-mining-peru/?fbclid=PAZXh0bgNhZW0CMTEAAacKmftget6cjyywiBRXgzdfR-GGQMJHPbyFKlW8Tmv44YlLDgZo7pA751D7ug_aem_ul4KL8pztvd0S4z-RSvGvA}{The methodology of MAPE} involves a compensation strategy based on replanting trees in areas that have lost forest cover because of mining and the use of optimisation of technologies to reduce the environmental impact of gold mining, especially the use of gravimetric tables, which mitigate partially the environmental impact of gold mining. The stated aim is also to formalise the value chain so that gold can be produced, partly mitigating the environmental impact, and the sale can be traceable with standards.

The interviews highlight how these women view mining as part of the region's future and take pride in this heritage, revealing a place-based narrative centred on gold. The interviewees were all from mining families that had lived in the region for generations. They were aware of the negative impact of mining on the environment and understood that this would have a detrimental effect on future generations. This appropriation of the \textit{“territory”} made them more open to promote activities of compensation for the environmental impact of their activities. They see the mining as part of the future of the region but one that embraces new technologies to mitigate environmental damage. 

Nevertheless, there were significant concerns amongst interviewees about the inadequacy of supporting institutions. The view is that informal mining (defined as mining within the legal concession area but not by the official title holder) has shot up because of the absence of monitoring and corruption in the police and army, extortion and insecurity. The feeling is that \textit{"the official laws that are enacted to change mining practices to support sustainability are very difficult for most miners to follow and so are ignored"}. Importantly, the success of these initiatives depends on the sale price of gold and the proof of origin certification. Yet, this type of practice for buying gold is incipient, especially in areas such as Bolivia where there are less controls. Therefore, regional actors have limited control over what happens in the value chain.

We identify MAPE as a placed-based initiative because it opens opportunities to experiment with alternative mining practices in the region and is driven by a deep concern for the future of the territory. It is noteworthy that, when asked, many of the women miners said that \textit{"they also owned land on which they practised agriculture"}. This brought them into the networks of farmers and their discussion over the introduction of more sustainable agricultural techniques. Participation in these Amazonian city networks helped to break down barriers between communities of producers and develop common identities to improving the future of the territory beyond their own personal interests.

Therefore, mining cooperatives might play a significant role in the governance of the territory, particularly in regions undergoing socio-environmental transformation. Although these actors are often associated with high and lasting environmental impacts, especially in fragile ecosystems such as the Amazon, their exclusion from environmental governance frameworks can lead to resistance and undermine the legitimacy of sustainability initiatives. Engaging place-based mining cooperatives in governance processes does not imply legitimising environmentally harmful practices, but rather acknowledges their structural role and embeddedness in territorial dynamics. Without their participation, environmental policies risk being contested or ignored, making the implementation of sustainable development projects significantly more challenging. A more inclusive governance approach, which acknowledges both the environmental risks and the socio-political role of these actors, may open up more viable pathways toward long-term socio-ecological and socio-technical transformation.

\subsection{Amazon Nut Collectors}

Amazon nut collectors represent a significant productive activity in this region. This initiative depends on the late stages of ecological succession for the natural production of Amazon Nuts. It is one of the most important local value chains for the sector and there are numerous local entrepreneurs and merchants involved in processing and selling Amazon nut oil and health derivates in MdD and beyond. The producers interviewed are organised under the Amazonian Chestnut Pickers Association from the Madre de Dios region (RONAP).

The concession covers an area of 32.000 ha of forest which are administered by 65 collectors. RONAP was formed to improve the market conditions for collectors in terms of prices, to introduce the children and families of the members into the sector, to provide technical help for members and to run an emergency health fund for members. The profits of the business are shared amongst the members. Crucially, it is also an activity that by its very nature benefits from the fruits and products of the forest – in others words the future of the business is directly linked to the health of the forest and upholding its regenerative features. RONAP works closely with international organisations such as RAINFOREST ALLIANCE, Sierra y Selva Exportadora, Conservación Amazónica and REDD+. 

From an ecology perspective, the nature of this activity makes it regenerative by nature, as collectors are trained to identify and help restore and maintain healthy ecosystems. RONAP also helps to formalise the process of nuts collection around a set of good practices that are recognised in the value chains and the international markets through the development of marketing tools such as QR codes that highlight the value of their work. They also process the nuts to generate added-value products which can be commercialise at the local and regional level. However, there also co-exists an informal market in Brazil and Bolivia. 

\subsection{Sintropic Agriculture}

The final example we identify is of Sintropic forestry or succession forestry. This alternative is important in contributing to accelerating the ecological succession of degraded forests. This is a method of farming that mimics natural ecological succession and stratification. It focuses on creating diverse and dynamic ecosystems that regenerate degraded soils, promote biodiversity and achieve high yields. It relies on natural processes to regulate the ecosystem, such as plant interactions, soil organisms, and weather patterns. The government rewards this type of agriculture by providing tax exemptions. 

The most successful of these is a 50ha farm called \textit{“Ten Paciencia”} (Be patient), 15 minutes by car outside the city of Puerto Maldonado. The origin of this farm goes back to 1990 when it was an intensive beef farm, at one point with 1200 Nelore cattle. Over the last 10 years, the number of cattle was reduced to 55 and the remaining cattle use regenerative techniques producing milk and cheese for local markets. 

Most of the land was then passed over to Sintropic agriculture, the principles of which are \textit{“density}, diversity, succession planting and stratification” which allows it use circular economy principles. Most of the sales are within the region – the owner mentioned that numerous attempts have been made to supply national markets in large cities, such as Cuzco, but most of these have been rejected to avoid becoming dependent on large-scale mono-production or on a single buyer. Instead, efforts have been made to support the business through eco-tourism, to support academic research projects and to provide training to agricultural farmers in the area through extension services. 

Today the farm produces 23 different amazonian fruits (zero chemicals) which are transformed into pulps, jams, biscuits, nectars, coffee, chocolates and soon wine. Also produced on site are 126 types of Amazonian plants. The fruits and plants are turned into dry fruits, medicines, forest products and there are bee hives to assist with pollination. 

\subsection{Synthesis of Madre de Dios examples}

Table 2 summarises the main results of the study of Madre de Dios. We identify that strategies of natural capital, such as compensation for protection of forests, inclusion of producers in international value chains and restoration of damaged forests, dominate sustainability efforts in the region. The implications for producers are significant. In most cases, these are further consolidated into international value chains, limiting their autonomy and scope of alternative actions. In addition, as actors in these value chains collaborate to formalise their activities, they attempt to negotiate their position within the governance of land use. Although these often have limited success given their position as small producers, actors open up these spaces to pressure for broader demands for local change that can be heard by organisation such as NGOs and local politicians.


\begin{center}
    \textbf{Table 2.} Initiatives in Madre de Dios
    
    \vspace{0.5em}
    \includegraphics[width=\textwidth]{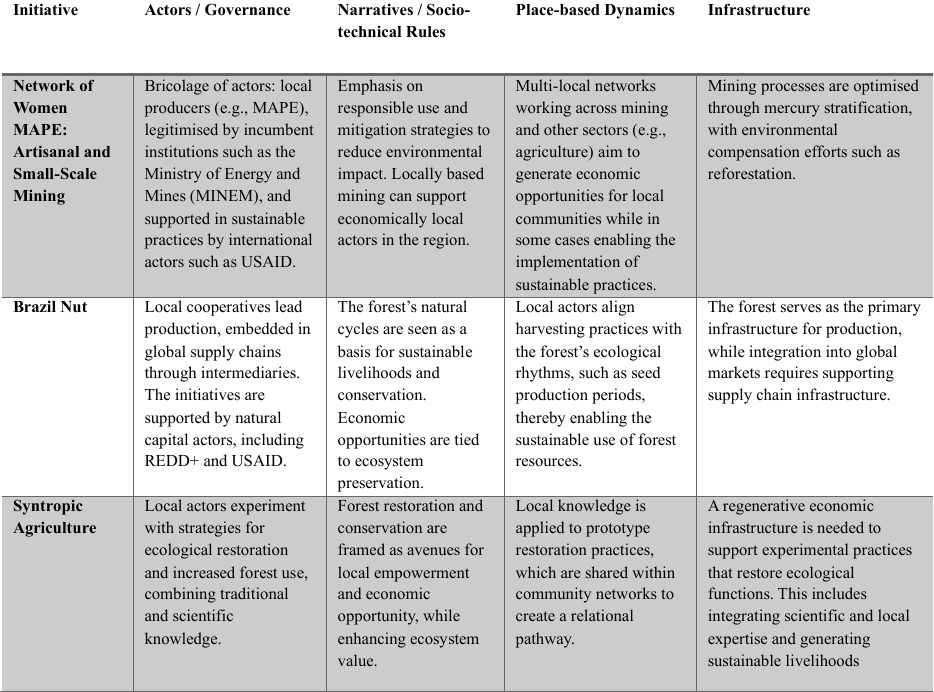}
\end{center}


\justifying

These initiatives represent an important step forward in a hitherto highly fragmented space with negligible institutions. Whilst the stated aim for this participation is often perceived as benefitting from funding opportunities, it is also clear that this is also about creating conditions for a \textit{“successful”} region where questions of environmental sustainability and how to confront illegal practices that harm the environment are discussed alongside questions of public safety, quality of health services and education. Through this process actors will continually re-define their territory. These cases, therefore, also highlight the importance of urban centres as places that provide a social infrastructure (universities, hotels, culture), particularly where institutions are still embryonic.

Nevertheless, a common feature of our interviews was also the continued difficulty of making a success of the formal economy and the fragile nature of its activities. These activities have to co-exist alongside a considerably larger number of informal, uncertified and at times illegal practices that pay no taxes and have far fewer overheads. In the case of the Amazon nut collectors in particular, the fragile nature of livelihoods is noticeable. Despite the fact that these communities are critical stewards of healthy forests, their ecological contribution as part of their labour is largely unrecognised or legitimised.

A local strategy that has emerged to try and achieve greater recognition for their activities and lower risks has been to construct a \textit{“commons-type institutions”} to support associations of producers engaged in formalisation of activity associated to reduction of deforestation. The most important of these has been the setting up of \textit{“La Marca Madre de Dios”} in 2019. This is a non-profit making territorial promotion entity that coordinates actions through public-private-civil partnerships. Officially, it has four guiding principles: safeguarding local biodiversity and the environment; modernize activities through professionalization, ethics and innovation of sustainable practices; contribute to sustainability and good living; and to promote the cultural identity of Madre de Dios. The process by which the Marca was formed emerged 2019 with the support of the French NGO (AFD) and several Peruvian institutes. The process took two years and was highly participatory with surveys and workshops the included miners, women and Indigenous groups. All three producers described in our cases are or have been \textit{“ambassadors”} of the Marca. La Marca and other similar initiatives are critical in providing common spaces in which to exchange information, compare best practices, establish a basis for collective action and search for common goals and a regional identity to its participants. 

\section{Discussion}

Unlike previous research in the transformation of deforestation, which has predominantly concentrated on identifying the drivers of deforestation \citep{lapola_drivers_2023, haddad_economic_2024, leclere_bending_2020}, our study provides a critical examination of potential pathways to reverse biodiversity loss, with particular attention placed on the role of local actors and their strategies in place-based dynamics. By analysing the institutional, technological infrastructures, and governance configurations that underpin deforestation in regions such as the Amazon, we offer insights into the underlying uncertainties and systemic complexities that constrain transformations in the context of biodiversity. In doing so, our findings contribute to broader debates on transformative pathways aimed at accelerating and operationalising sustainability transitions in related contexts.

\citet{stirling_keep_2010, stirling_direction_2009, stirling_general_2007}
emphasises that oversimplifying complex and interrelated socio-ecological challenges often leads to conflict escalation and short-term solutions. In this direction, our findings highlight the importance of embracing plural and diverse perspectives in addressing deforestation in the Amazon. Navigating uncertainty and complexity requires inclusive deliberation \citep{arroyave_social_2021}, particularly when pursuing deep structural institutional changes. For example, our results suggest that the effectiveness of the natural capital pathway has been constrained by poor governance and limited efforts to construct local institutions that would allow these initiatives to expand. These are essential for identifying trade-offs, mitigating blind spots, and enhancing the long-term resilience of sustainability strategies \citep{harvey_spaces_2008}. Therefore, the deliberate inclusion of multiple forms of knowledge and local agency should be considered a critical condition for the viability and legitimacy of any attempt to transform the deforestation regime.

In addition, we argue that place-based coalitions can be critical for addressing interconnected social and environmental challenges and facilitating the emergence of regenerative practice. The historical trajectory of deforestation in the Amazon reveals that collaborative efforts among diverse local actors created key opportunities in the 1990s to reframe the use of the forest and confront the social injustices embedded in the early phases of the deforestation regime. 

In a similar way to what \citet{ramirez_fostering_2020} found in relation to a study of wetlands defence, bricolage processes \citep{karnoe_path_2012, boschma_towards_2017}, whereby actors recombine available resources, knowledge, and institutional arrangements to prototype context-specific solutions can be relevant for Amazon communities. In that case, the co-location scientists and grassroots movements forged durable alliances supported by legal instruments. In the case of Madre de Dios, Peru, a range of actors including local associations, local intermediaries and NGOs operating within a highly decentralised and weakly regulated governance context may play this role. This fragmented institutional landscape underscores the structural barriers to scaling and institutionalising grassroots experimentation in settings marked by policy volatility, insecure land tenure, and the pervasive influence of global market logics. In these contexts some place-based strategies would seem essential to build the bases for institutionalisation of regenerative change alternatives. 

\citet{dasgupta_economics_2021} highlights negative feedback loops within the natural capital practices— particularly those treating nature as a financial surplus—can both enable and constrain opportunities for change. Instruments such as Payments for Ecosystem Services (PES), biodiversity credits, and carbon trading schemes reflect this logic by attempting to internalise ecological externalities and incentivise conservation through market mechanisms. However, in practice, systemic market failures and the absence of bottom-up experimentation may undermine the transformative potential of these approaches.

In addition, the natural capital approach appears to be increasingly trapped in a trajectory of greenwashing, failing to effectively address biodiversity loss and exacerbating existing social inequalities. Thus, while the natural capital approach introduces new actors and valuation instruments, it frequently overlooks local knowledge systems, entrenched historical injustices, and the deep contextual uncertainties that shape deforestation dynamics. Such technocratic oversimplifications can reinforce power asymmetries, escalate conflict, and result in short-term or maladaptive interventions. Therefore, our results suggest that the reconfiguration framework on its own, expressed in the natural capital, may fall short in building opportunities to generate transformative change in the context of biodiversity

\section{Conclusion}

The findings of this study addressed the overall question: What types of sustainability pathways might weaken the system of deforestation in the Amazon, and under what conditions can they emerge and evolve? This question has been addressed by identifying and contrasting two main pathways—natural capital associated with optimisation and reconfiguration, and regenerative change. In responding to this question and in contrast to previous pathways studies in transitions studies, the paper incorporated not only time but also space, which allowed us to see how the historical and institutional analysis of the Amazon – including its collective memory, identities and human-related practices are deeply related to its natural and ecological system and define a relational interaction with the territory. 

Our findings suggest that diverse actors are actively experimenting with hybrid approaches that draw on elements from natural capital and regenerative change. In the second half of the last century included infrastructural expansion and extractivist land-use narratives that have stabilised a regime of deforestation. However, they also include broad-based alliances between local counter movements and international organisations that have contested these extractivist logics, although some attempts to reform it through market-based instruments have frequently reproduced the same logics they seek to replace.

A central feature of our argument has been that as part of the contestation process the construction of a regenerative alternative pathways to end or significantly limit deforestation will depend to a large degree on the strategies of the communities for dignity, recognition and acknowledgment of their rights. These can express themselves in efforts to undo certain injustices. We also saw that some seek pragmatic solutions that work with sustainability narratives.

The central point here is that these struggles and strategies need to be internalized as part of any sustainability pathways. Our concern is that the reconfiguration framework, which relies primarily on regulative changes and financial incentives, may have short-term effects but lack the resilience if not linked to the wider social transformations in the territories. The regenerative change pathways try to do this by combining ecological practices, community conservation and local knowledge systems. But this is a difficult pathway to construct without significant investment and support. The likely scenario is that grassroots actors will respond to multiple elements from the three pathways. 

Finally, while this study strives to provide both theoretical and empirical insights, some limitations should be acknowledged. The empirical evidence is concentrated in the Madre de Dios region of Peru, which—although highly relevant—is not fully representative of the diverse socio-political and ecological conditions across the broader Amazon basin. A comparative study might provide better details about the diverse approaches to address deforestation. Additionally, while our interviews and literature review offer rich insights into emergent place-based strategies, further quantitative research would be valuable to strengthen and validate the findings. Yet, this paper contributes to ongoing debates on sustainability transitions in biodiverse environments by showing that context-sensitive, participatory, and historically grounded approaches are essential for building more inclusive and resilient pathways. It also underlines the importance of engaging more deeply with how transitions unfold in places marked by institutional instability and in communities that have been historically affected by long-term injustices.

\section*{CRediT authorship contribution statement}

\textbf{Oscar Yandy Romero Goyeneche:} Conceptualisation, Literature review, writing – original draft. \textbf{Matias Ramirez:} Conceptualisation, 
writing – original draft, data collection and interviews analysis. \textbf{Ana Milena Osorio-Garcia:} Conceptualisation, Data collection and interviews analysis. \textbf{Ursula Harman Canalle:} Conceptualisation, Data collection and analysis.

\section*{Declaration of competing interest}

The authors declare that they have no known competing financial interests or personal relationships that could have appeared to influence the work reported in this paper.

\section*{Acknowledgements}

We gratefully acknowledge the support of the Wyss Academy for Nature and the Transformative Innovation Policy Consortium for providing the funding that made this research possible. We also thank the Antioquia Technology Center (CTA) for its support in administering the project. Special thanks to Csaba Földesi, Rob Alkemade, Jeanne Nel, and the entire Hub on Bending the Curve for Biodiversity Loss for their valuable discussions, inspiration, and support during the early phase of this research. Lastly, this work would not have been possible without the kind support of the Artisan Women Gold Miners (MAPE), Amazonian Chestnut Pickers Association from the Madre de Dios region (RONAP), and Salomón Perez Alencart.

\bibliographystyle{unsrtnat}  

\end{document}